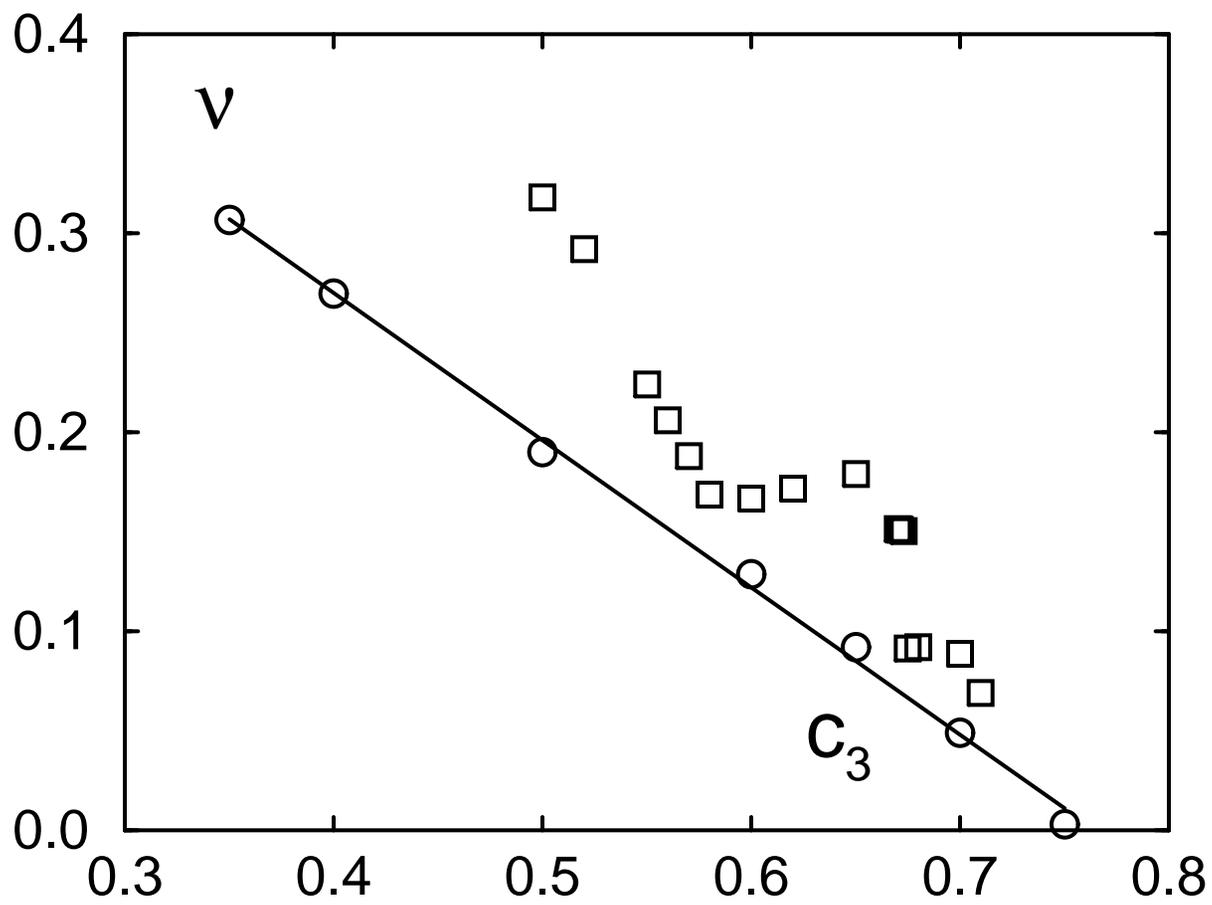

Fig.1 Torcini

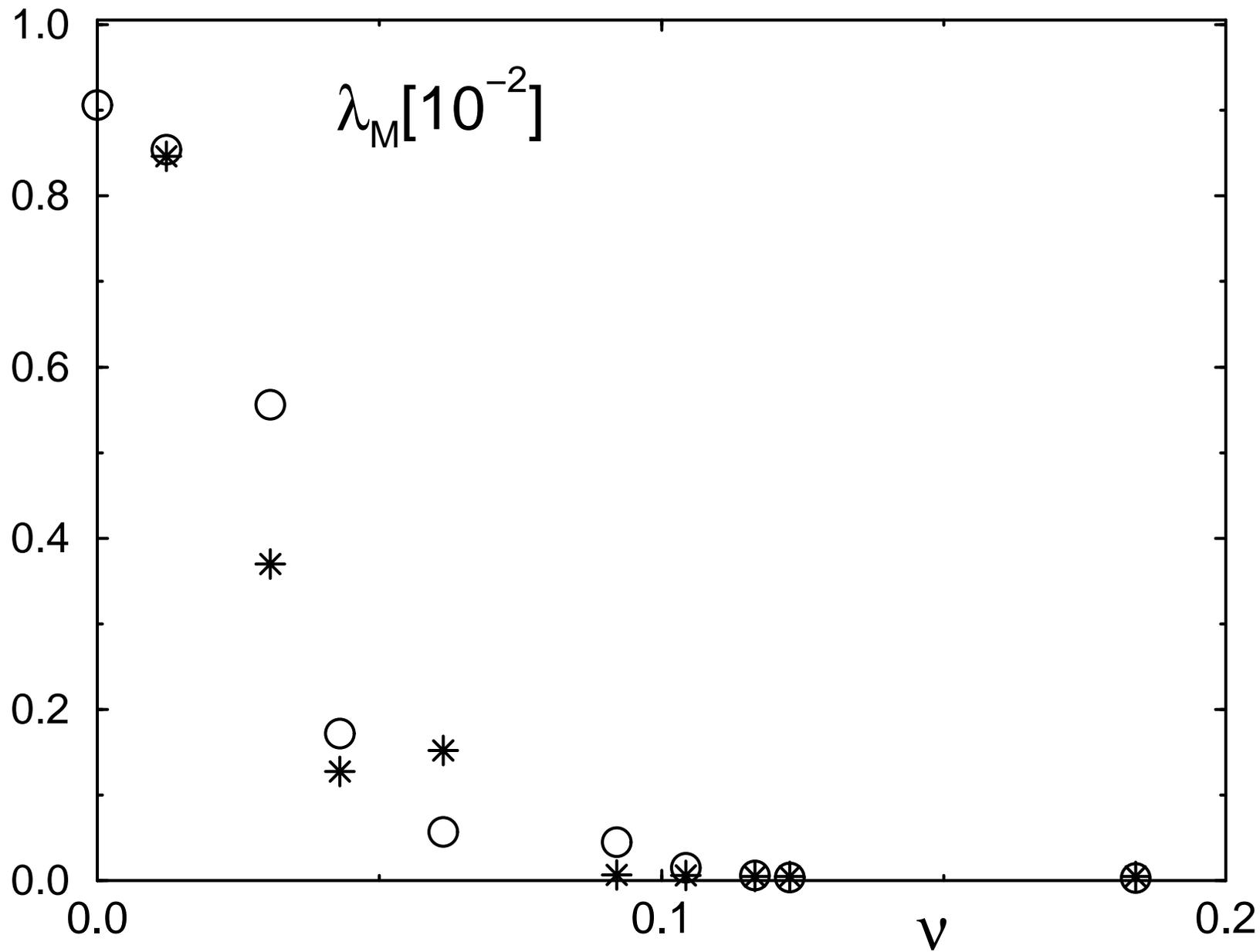

$\lambda_M[10^{-2}]$

$\nu$

Fig.2 Torcini

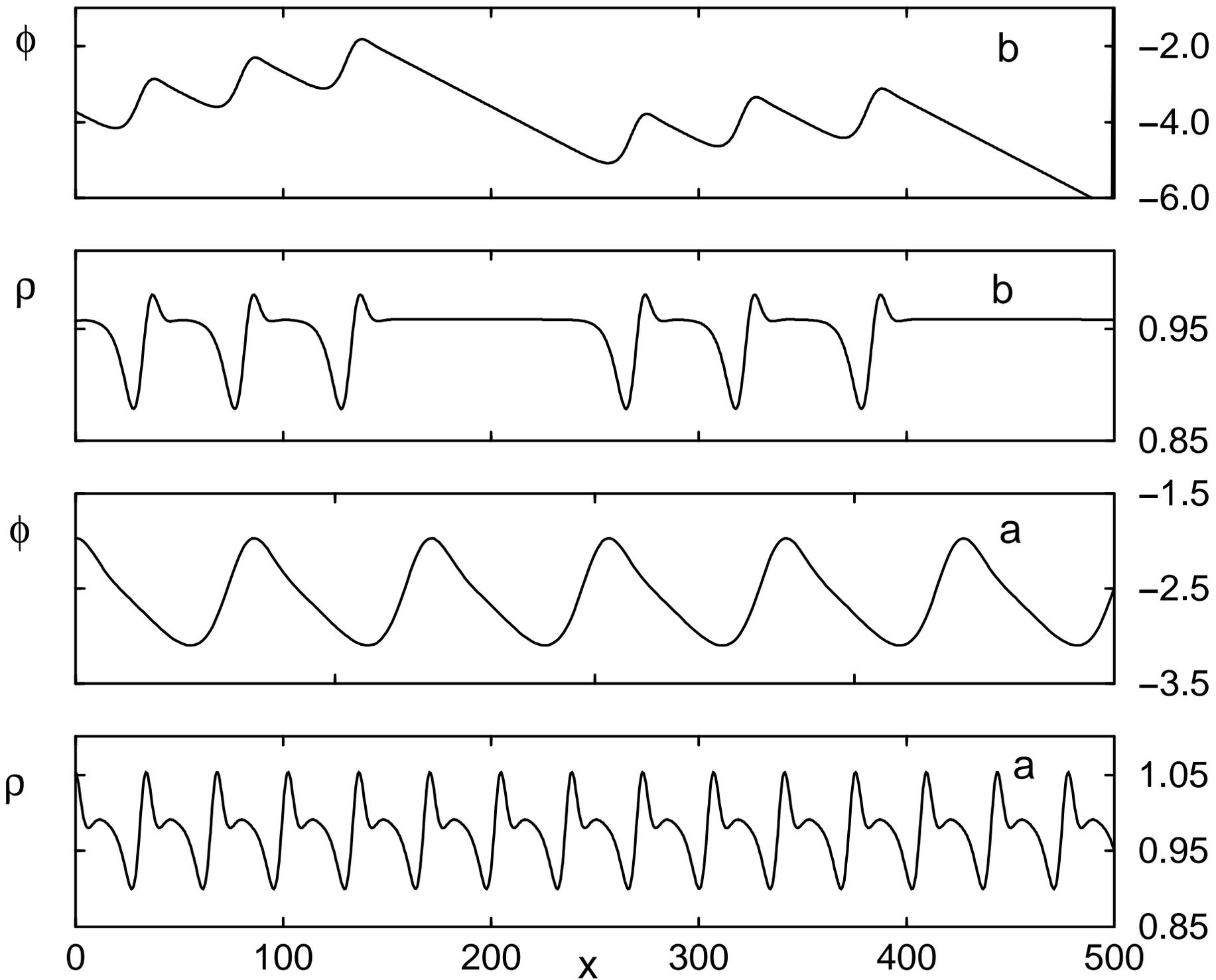

# Order Parameter for

# the Transition from Phase to Amplitude Turbulence

Alessandro Torcini

*Theoretical Physics, Wuppertal University, D-42097 Wuppertal, Germany*

## Abstract

The maximal conserved phase gradient is introduced as an order parameter to characterize the transition from phase- to defect-turbulence in the complex Ginzburg-Landau equation. It has a finite value in the phase-turbulent regime and decreases to zero when the transition to defect-turbulence is approached. Solutions with a non-zero phase gradient are studied via a Lyapunov analysis. The degree of "chaoticity" decreases for increasing values of the phase gradient and finally leads to stable travelling wave solutions. A modified Kuramoto-Sivashinsky equation for the phase-dynamics is able to reproduce the main features of the stable waves and to explain their origin.

Typeset using REVTₑX

Spatially extended chaotic systems have been recently the subject of several theoretical and experimental investigations [1]. In spite of remarkable progresses, a still open problem concerns the possibility to describe nonequilibrium phase transitions with concepts borrowed from statistical mechanics [1–4]. The complex Ginzburg-Landau equation (CGLE) is one of the most appropriate model to study such transitions, because is universal [1,5] and experimentally relevant [6]: the dynamics of extended systems undergoing a Hopf bifurcation from a stationary to an oscillatory state is described by the CGLE [7]. Moreover, several physical, chemical and biological phenomena can be well reproduced through the CGLE [1]. Even in one dimension the CGLE displays a variety of dynamical regimes and phase transitions [2–4,7–9]. Among them much attention has been devoted to the transition between two different type of chaotic phases [2–4,7,8]: namely, the phase-turbulent (PT) and the defect-turbulent (DT) regimes. However, this transition has been mainly studied from the DT side [2–4].

This Letter will be focused on the PT regime and in particular on the introduction of an order parameter for the above mentioned transition. To be more specific, let me write the one-dimensional CGLE as

$$A_t = (1 + ic_1)A_{xx} + A - (1 - ic_3)|A|^2 A \tag{1}$$

where the parameters $c_1$ and $c_3$ are real positive numbers, and $A(x,t) = \rho(x,t)\exp[i\psi(x,t)]$ is a complex field of amplitude $\rho$ and phase $\psi$. Of particular interest in the $(c_1, c_3)$-plane is the so-called Benjamin-Feir line (BFL) defined by $c_3 = 1/c_1$, which identifies the linear stability limit for the plane-wave solutions of (1) [5]. The PT regime is encountered just above the BFL (i.e. for $c_3 > 1/c_1$) [7]. In this state the chaotic behaviour of the field is essentially ruled by the dynamics of the phase. Moreover, the amplitude is always bounded away from zero, accordingly; the average phase gradient $\nu = \frac{1}{L}\int_0^L dx\,\partial_x\psi(x,t)$ is conserved for periodic boundary conditions [7]. Another important line $(L_1)$ separates the PT state from a more chaotic phase: the DT regime. In this state, amplitude-dynamics becomes predominant over phase-dynamics [2,7,8]. In particular, large amplitude oscillations are observed, occasionally driving $\rho(x,t)$ to zero, in which case the phase is no longer well defined and $\nu$ is not conserved. The vanishing of the amplitude determines a so-called space-time defect. The density of defects $\delta_D$ is a good order-parameter to characterize the transition from the DT to the PT phase; in fact its value is $> 0$ in the DT regime, while it vanishes approaching the PT phase [2,4]. Conversely, none of the other parameters introduced so far (e.g. phase and amplitude correlations lengths, or the Kaplan-Yorke dimension density) reveals a clear signature of the transition [3,4].

In the PT regime, a relevant parameter to characterize a state of the CGLE is the value of $\nu$, which is a conserved quantity in the absence of defects. For each value of $\nu < 1$ there exists a plane-wave solution of the CGLE

$$A(x,t) = \sqrt{1-\nu^2}\exp[i(\nu x + \Omega_0 t)] \tag{2}$$

where $\Omega_0 = c_3 - (c_1 + c_3)\nu^2$. Below the BFL, these solutions are stable against long-wavelength instabilities for $\nu^2 < (1 - c_1 c_3)/(2(1 + c_3^2) + 1 - c_1 c_3)$ [5], while they turn out to be all unstable in the PT regime. However, it is reasonable to expect that different $\nu$ values will characterize distinct classes of chaotic or quasi-periodic solutions also in the PT phase. Nevertheless, the majority of analyses reported in literature has been devoted to solutions with $\nu = 0$.

In this Letter, I determine the maximal value $\nu_M$ of the conserved phase gradient in statistically stationary states, i.e. in the limit $L \to \infty$ and $t \to \infty$. I shall argue that this can be used as an order parameter, with $\nu_M = 0$ in the DT regime, with $\nu_M > 0$ in the PT state, and with a smooth change at the transition line $L_1$. Moreover, a Lyapunov analysis reveals that for increasing $\nu$-values the solutions are less and less chaotic. This behaviour can be explained assuming that the phase-dynamics is ruled by a modified Kuramoto-Sivashinsky equation (MKSE) [7,10–12]. For reasonably high $\nu$-values travelling pulse-train solutions are found to be stable. The main parameters characterizing such solutions can be derived from the above mentioned MKSE.

The analysis here reported is limited to a parameter region where it is known that the CGLE shows a continuous transition from the PT to the DT regime [2,4], namely $c_1 = 3.5$ and $c_3 \geq 1/c_1$. The integration scheme here adopted is a time splitting code where the integration of the spatial derivatives is not performed, as usual, in Fourier space, but instead is computed in $x$-space through a convolution integral [13,14]. The adopted integration parameters are: spatial resolution $dx = 0.5$; integration time step $dt = 0.05$; number of grid points $N = L/dx$ ranging from 2048 up to 8192, and periodic boundary conditions.

In order to study solutions with a non zero value of $\nu$, an initial state with $\nu \neq 0$ has been prepared and its evolution followed. Due to the absence of defects in the PT regime, $\nu$ should be conserved, at least for small initial values. I have numerically verified that $\nu$ is indeed conserved for all values below an upper limit $\nu_M$. For $\nu > \nu_M$ defects eventually arise leading, after a readjusting time, to a $\nu$-value smaller than $\nu_M$. This behaviour can be understood by observing that the minimal $\rho$-value decreases continuously for increasing $\nu$, until, for $\nu > \nu_M$, the amplitude can eventually vanish with the consequent emergence of defects. In order to evaluate the maximal phase gradient accurately, the conservation of each $\nu_M$ value reported in Fig. 1 has been checked considering from 50 to 70 different initial conditions. In particular, each configuration has been followed for a time $t \geq 10,000$, after a reasonably long transient had been discarded. Moreover, for each examined $c_3$-value, some trajectories have been followed for longer times, typically of the order of $t = 150,000$. No dependence of $\nu_M$ on $L$ has been observed, considering chain lengths from $L = 1024$ up to $L = 4096$. Fig. 1 shows a roughly linear decrease of $\nu_M$ for increasing values of $c_3$. A linear fit of the data gives $\nu_M \simeq -0.74c_3 + 0.57$. Assuming that the linear behaviour extends up to the $L_1$ line, a critical value $c_3^* = 0.76 \pm 0.03$ is obtained. This result is in reasonable agreement with the values obtained from the defect density reported in literature [2,4], as well as with our own measurements of $\delta_D$, which give $c_3^* = 0.755 \pm 0.002$ [13,15].

As a further characterization of these solutions, I have evaluated the maximal Lyapunov exponent $\lambda_M$ for several values of $c_3$ and $\nu$. I noticed multistability for $\nu > 0$, i.e. several coexisting attractors for each $\nu$ value, different attractors being characterized by different

Lyapunov exponents. However, a characteristic common to all these initial conditions considered is that $\lambda_M$ tends to decrease with increasing $\nu$, except for small fluctuations (see Fig. 2). In particular, for $c_3 \leq 0.5$ and for sufficiently high values of $\nu$, non-chaotic states are found. Two different kinds of non-chaotic states have been observed: the former (type $\alpha$) is spatially periodic with spatial wavevector $q = \nu$, formed of identical "pulse-like" structure of length $L_P = 2\pi/\nu$; the latter (type $\beta$) shows essentially periodic regions separated by domains in which $\rho$ is constant and the phase decreases linearly (see Fig. 3). The selection of these patterns depends on the initial conditions. Numerically I found that all non-chaotic solutions are of the form:

$$A(x,t) = h(x - vt)\mathrm{e}^{i(\nu x + \omega t)} \tag{3}$$

where $h(\xi) = \rho(\xi)\exp[i\psi_0(\xi)]$ is in general complex. Amplitude and phase can be written as

$$\rho(x,t) = \rho(\xi) \quad ; \quad \psi(x,t) = \psi_0(\xi) + \omega t + \nu x \tag{4}$$

with $\xi = x - vt$.

For spatially periodic patterns, the elementary pulses of length $L_P$ are stable solution of the CGLE for a short chain of length $L = L_P$ and with a phase gradient $\nu_P = 2\pi/L_P$, i.e. the minimal non-vanishing phase-gradient for such a periodic system. These travelling pulses originate through a bifurcation from the plane-wave solutions (2). Moreover, they are stabilized in short systems, because of the long-wavelength instability cutoff. A more extensive and detailed study of these short-chain solutions will be reported elsewhere. Here, I want just to point out that solutions with a conserved $\nu_P$ are no longer observable below a minimal length $L_{min}$. The corresponding phase gradient $\nu_U = 2\pi/L_{min}$ is an upper bound for $\nu_M$, as shown in Fig. 1. For increasing $c_3$, the value $\nu_M$ is better and better approximated by $\nu_U$. A naive explanation of this fact can be given by assuming that in the proximity of $\nu_M$, only stable solutions of type $\alpha$ are observed. For these solutions the minimal value of the spatial period is obviously $L_{min}$ and the maximal possible phase gradient $\nu_U$. However, in the limit $\nu \to \nu_M$, for $c_3 \le 0.5$ solutions of type $\alpha$ coexist with those of type $\beta$, while for $c_3 > 0.5$ slightly chaotic solutions, formed by an array of elementary pulses of different lenghts $L_p \ge L_{min}$, are observed. All these facts imply that $\nu_U \ge \nu_M$.

As a final point, I would like to explain why $\lambda_M$ decreases with increasing $\nu$. This can be done by recalling that just above BFL, the dynamics is essentially ruled by the phase behaviour, since the amplitude can be considered as a "slaved" variable of the phase [7,8,12]. A phase modulation $\phi(x,t) = \psi(x,t) - \nu x = \psi_0(\xi) + \omega t$ on the plane wave solution (2) satisfies a MKSE [11],

$$\dot{\phi} - \Omega_0 + \Omega_1 \partial_x \phi + \Omega_2^{(1)} \partial_x^2 \phi + \Omega_2^{(2)} (\partial_x \phi)^2 + \Omega_3 \partial_x^3 \phi + \Omega_4^{(1)} \partial_x^4 \phi + \Omega_4^{(2)} \partial_x \phi \partial_x^3 \phi = 0 \qquad (5)$$

where $\Omega_1 = 2\nu(c_1 + c_3)$, $\Omega_2^{(1)} = c_1 c_3 - 1$, $\Omega_2^{(2)} = (c_1 + c_3)$, $\Omega_3 = 2\nu c_1(1 + c_3^2)$, $\Omega_4^{(1)} = c_1^2(1 + c_3^2)/2$ and $\Omega_4^{(2)} = 2c_1(1 + c_3^2)$. In order to check the validity of Eq. (5) for the dynamics of the phase for the CGLE in the PT regime, I derive from Eq. (5) an expression for the velocity $v$ and the frequency $\omega$ of the travelling solutions (3). Following [11], I obtain

$$\omega = \Omega_0 - \Omega_2^{(2)} < (\partial_\xi \psi_0)^2 > + \Omega_4^{(2)} < (\partial_\xi^2 \psi_0)^2 > \qquad (6)$$

$$v = \Omega_1 + (\Omega_2^{(2)} < (\partial_\xi \psi_0)^3 > - \Omega_3 < (\partial_\xi^2 \psi_0)^2 > + \Omega_4^{(2)} < (\partial_\xi \psi_0)^2 \partial_\xi^3 \psi_0 >)/ < (\partial_\xi \psi_0)^2 > \qquad (7)$$

where $< \cdot >$ is the average along the chain and over several consecutive realizations. The two expressions can easily be obtained by noticing that for a solution of type (4) the temporal derivative of the phase can be written as $\dot{\phi} = \omega - v \partial_\xi \psi_0(\xi)$, while $\partial_x \phi(x) = \partial_\xi \psi_0(\xi)$. Substituting the temporal derivative into Eq. (5) and averaging both sides of the equation leads to Eq. (6). To obtain Eq. (7) both sides of Eq. (5) should be multiplied by $\partial_\xi \psi_0(\xi)$ before averaging. Several quantities appearing in Eq. (5) have zero average due to the periodic boundary conditions. Therefore the final expressions for $\omega$ and $v$ are drastically simplified. By inserting the phase values obtained from simulations of the CGLE into Eqs. (6) and (7), a very good agreement with the measured quantities is indeed achieved (see Table 1). Nevertheless, I expect that the phase description (5) will become less and less accurate approaching the $L_1$ line and will finally break down when the amplitude dynamics becomes predominant. However, I have verified that in the whole examined parameter interval the amplitude dynamics is essentially ruled by the phase, except when defects arise.

Assuming that Eq. (5) describes sufficiently well the dynamics of the CGLE-phase, let me now explain the observed non-chaotic behaviours. In Ref. [11] it has been shown that Eq. (5), rewritten for the variable $\partial_x\phi$, belongs to a class of equations that shows stable solutions formed by pulse-like periodic structures travelling along the chain with a finite velocity [16,17]. The birth of these non-chaotic solutions is due to the presence of the dispersion term $\propto \partial_x^3\phi$ [16]. In particular, in Ref. [17] it has been shown that this equation is multistable: chaotic and non chaotic solutions can coexist. However, stationary periodic attractors prevail over the strange ones for increasing values of the dispersion constant $\Omega_3$ and, above a threshold value, only non chaotic solutions are observed. Since $\Omega_3$ is proportional to $\nu$, this explains the decreasing behaviour of $\lambda_M$ for increasing $\nu$, as well as the origin of the observed travelling wave solutions for high $\nu$-values. Finally, it should be remarked that for $c_3 > 0.5$ it is no longer possible to observe stable solutions, because they would occur for value of $\nu > \nu_M$.

In conclusion, I emphasize that $\nu_M$ turns to be a good order parameter to describe the transition from the PT to the DT phase. The role of this new parameter is complementary to that of the defect density, since it is non-zero in the PT regime, while the defect density is non-zero in the DT state. Moreover, the dynamics in the PT phase depends strongly on the $\nu$-value: for $\nu \to 0$ a predominance of chaotic solutions is observed, while for $\nu \to \nu_M$ non-chaotic behaviours prevail and a new family of stationary solutions emerges.

## ACKNOWLEDGMENTS

I am indebted with P. Grassberger for extremely effective suggestions. I want also to thank H. Frauenkron, S. Lepri and A. Politi for helpful discussions. Supports from the European Community under the grant no ERBCHBICT941569 and from the Cooperativa Fontenuova are also gratefully aknowledged.

than that required for a FFT code by a factor $\simeq 1/\ln(N)$. In my applications, this factor was $\simeq 0.3$. For more details see [13].

<center>FIGURES</center>

**Fig.1**: Maximal phase gradient $\nu_M$ as a function of the parameter $c_3$ (circle). The values for $\nu_U$ are also shown (squares). The solid line represent a linear fit for the $\nu_M$ data.

**Fig.2**: Maximal Lyapunov exponents versus the phase gradient $\nu$ for two different sets of initial conditions for $c_3 = 0.5$. The two initial conditions correspond to states with phase gradient $\nu$ and with noise added only on the amplitude (circle) or added on both amplitude and phase (asterisks). The data have been obtained for system size $L = 1024$ and for integration times $t = 130,000 - 250,000$.

**Fig.3**: Phase and amplitude for the two kinds of observed non chaotic solutions: a) solution of type $\alpha$ with $c_3 = 0.5$ and $\nu = 0.184$; b) solution of type $\beta$ with $c_3 = 0.35$ and $\nu = 0.307$.

TABLES

TABLE I. The parameters characterizing the stable solutions (3), namely the propagation velocity $v$ and the frequency $\omega$, are reported for various $\nu$ values. The values are estimated using the expressions (6) and (7). In parentheses are the values measured directly from the travelling waves solutions. The data in the table refer to $c_3 = 0.5$, except for $(*)$ that is for $c_3 = 0.4$. The label $(+)$ indicates solutions corresponding to different initial conditions, but with the same $\nu$.

| $\nu$ | $\omega$ | $v$ | $\nu$ | $\omega$ | $v$ |
|---|---|---|---|---|---|
| 0.123 | 0.428 (0.427) | 1.26 (1.24) | $0.123^{(+)}$ | 0.428 (0.424) | 1.06 (1.09) |
| 0.184 | 0.355 (0.341) | 1.59 (1.65) | $0.184^{(+)}$ | 0.344 (0.338) | 1.58 (1.63) |
| 0.245 | 0.229 (0.223) | 1.93 (1.98) | $0.282^{(*)}$ | 0.074 (0.070) | 2.37 (2.34) |